\documentclass[twocolumn]{svjour3}

\usepackage{hyperref}
\usepackage{xcolor}
\usepackage{amsmath}
\usepackage{amssymb}
\usepackage{amsxtra}
\usepackage[utf8]{inputenc}
\usepackage[mathscr]{eucal}
\usepackage[english]{babel}

\journalname{NoDy}

\usepackage{cite}
\usepackage{graphicx} 
\usepackage{setspace}

\hypersetup{
    pdftitle = {Non-stationary localized oscillations},
}

\def\d{\mathrm d}

\def\prpr#1{#1^{\prime\prime}}

\def\Int{\int_{-\infty}^{+\infty}}

\def\const{\operatorname {const}}
\def\sign{\operatorname {sign}}

\def\SQRT{\sqrt{1-\Omega^2}}

\def\sign{\operatorname{sign}}
\def\const{\operatorname{const}}

\def\W{{W_0}}

\def\tauu{\tau}
\def\I{\mathrm i}
\def\NEW#1{{\color{black}#1}}

\journalname{NoDy}
\title{Non-stationary localized oscillations of an infinite string,
with time-varying tension, lying on the
Winkler foundation with a point elastic inhomogeneity}
\titlerunning{Non-stationary localized oscillations of an infinite string 
with time-varying tension} 
\author{S.N.~Gavrilov \and E.V.~Shishkina \and Yu.A.~Mochalova}

\institute{
S.N.~Gavrilov, corresponding author
\at
Institute for Problems in Mechanical Engineering RAS, V.O., Bolshoy pr.~61,
St.~Petersburg, 199178, Russia. \\
\email{serge@pdmi.ras.ru}           
\and
S.N.~Gavrilov 
\at
Peter the Great St.~Petersburg Polytechnic University (SPbPU),
Polytechnicheskaya str.~29, St.~Petersburg, 195251, Russia.
\and
E.V.~Shishkina 
\at
Institute for Problems in Mechanical Engineering RAS, V.O., Bolshoy pr.~61,
St.~Petersburg, 199178, Russia. \\
\email{shishkina\_k@mail.ru}
\and
Yu.A.~Mochalova
\at
Institute for Problems in Mechanical Engineering RAS, V.O., Bolshoy pr.~61,
St.~Petersburg, 199178, Russia.\\
\email{yumochalova@yandex.ru}}

\begin{document}
\maketitle

\begin{abstract}
We consider non-stationary oscillations of an infinite string with
time-varying tension. The string lies on the Winkler foundation with a
point inhomogeneity (a concentrated spring of negative
stiffness).  In such a system with constant parameters (the string tension), under certain conditions a trapped mode of oscillation exists and is unique. Therefore, applying
a non-stationary external excitation to this system
can lead to the emergence of the string oscillations localized near the inhomogeneity.
We provide an analytical description of non-stationary localized oscillations 
of the string with slowly 
time-varying tension
using the asymptotic procedure based on successive application of two asymptotic
methods, namely the method of stationary phase and the method of multiple
scales. The obtained analytical results were verified by
independent numerical calculations based on the finite difference method.
The applicability of the analytical formulas was
demonstrated for various types of external excitation and laws governing the
varying tension. In particular, we have shown that 
in the case when the trapped mode frequency approaches zero, localized low-frequency oscillations with increasing
amplitude precede the localized string buckling.  The dependence of the amplitude of
such oscillations on its frequency is more complicated in comparison with the
case of a one degree of freedom system with time-varying stiffness.
\end{abstract}

\section{Introduction}


In this paper we consider a mechanical system with mixed spectrum of natural
oscillations. Namely, we deal with an infinite string 
with slowly time-varying tension. The string lies
on the Winkler foundation with a
point inhomogeneity (a concentrated spring of negative
stiffness).  In the case
of a constant string tension the discrete part of the spectrum for such a
system may contain
unique (positive) eigenvalue, which is less than the lowest
frequency for the string on the uniform foundation. This
special natural frequency corresponds to a trapped mode of oscillation 
with eigenform localized near the spring. The phenomenon of
trapped modes was discovered in the theory of surface water waves
\cite{ursell1951trapping}. The examples of various mechanical systems, where
trapped modes can exist, can be found in studies
\cite{kaplunov1986torsional,
abramyan1992characteristics,
kaplunov1995simple,
abramyan1998trapping,
Gavrilov-2006-trans,
gavrilov2002etm,
alekseev2002vibration,
mciver2003excitation,
indeitsev2004localization,
porter2007trapped,
kaplunov2008example,
motygin2008trapping,
nazarov2010sufficient,
pagneux2013trapped,
porter2014trapped,
gavrilov2016trapped,
kaplunov2005localized,
Ind-book-R2E,
indeitsev2000resonance,
indeitsev2012motion,
wang2014vibration,
indeitsev2015localization,
indeitsev2016evolution,
gavrilov2017trapped,
shishkina2018non}. 

It is known \cite{kaplunov1986torsional,gavrilov2002etm,mciver2003excitation,
Ind-book-R2E,
indeitsev2016evolution,
gavrilov2017trapped,shishkina2018non}
that applying non-stationary external excitation to a system
possessing trapped modes leads to the emergence of undamped oscillations
localized near the inhomogeneity. The large time asymptotics for such oscillation
can be found 
\cite{gavrilov2002etm,mciver2003excitation,
indeitsev2016evolution,
gavrilov2017trapped,shishkina2018non}
by means of the method of stationary phase \cite{fedoruk1977,nayfeh1993}. 
Gavrilov in \cite{gavrilov2002etm,Gavrilov-2006-trans} 
suggested an asymptotic procedure based on successive application of two asymptotic
methods, namely the method of stationary phase \cite{fedoruk1977,nayfeh1993} 
and the method of multiple scales
\cite{nayfehperturbation,nayfeh1993} that allows us to investigate
non-stationary processes in perturbed systems, with slowly time-varying
parameters,
possessing trapped modes. In 
studies 
\cite{gavrilov2002etm,Gavrilov-2006-trans}
the problem concerning non-uniform motion of a point mass along a taut string
on the Winkler foundation was considered and solved. Note that later the same problem was
reconsidered in paper 
\cite{gao2014exact} by 
Gao, Zhang, Zhang, and Zhong
in very particular case of uniform motion at a given speed. 

The asymptotic procedure suggested in 
\cite{gavrilov2002etm,Gavrilov-2006-trans} was successfully applied to
describe the evolution of the amplitude of the trapped mode of oscillations in a
taut string on the Winkler foundation with a point inertial inclusion of
time-varying mass \cite{indeitsev2016evolution} and in a taut string on the Winkler foundation
with a concentrated spring of negative time-varying stiffness
\cite{gavrilov2017trapped}. All problems considered in previous papers 
\cite{gavrilov2002etm,Gavrilov-2006-trans,indeitsev2016evolution,gavrilov2017trapped} 
are formulated for the Klein-Gordon PDE with constant coefficients.
In our recent paper \cite{shishkina2018non} a Bernoulli-Euler beam on the Winkler foundation
with a concentrated spring of negative time-varying stiffness
is considered. In the latter case, a PDE with
constant coefficients is also under consideration.
This allows one to verify the obtained analytical results by reduction of
the problem to a Volterra integral equation of the second kind with its kernel
expressed in terms of the fundamental solution for the corresponding PDE. 
The integral equation can be easily solved numerically. This was done in all
previous studies
\cite{gavrilov2002etm,Gavrilov-2006-trans,indeitsev2016evolution,gavrilov2017trapped,shishkina2018non}, and a  good mutual agreement between asymptotic and
numeric results was demonstrated. 

The consideration of the problem for a string with variable tension has some
geophysical motivation \cite{gavrilov2016trapped}. In contrast with 
previous papers
\cite{gavrilov2002etm,Gavrilov-2006-trans,indeitsev2016evolution,gavrilov2017trapped},
the governing equation here is the
Klein-Gordon equation with time-varying coefficients. To verify the
obtained analytical results we solve numerically the initial value
problem for this PDE 
using the finite difference method 
and demonstrate the applicability
of the analytical formulas for various types of external excitation and 
laws governing the varying tension.
Some preliminary results concerning this problem were
presented in a recent conference publication \cite{gavrilov2017trapped}. 

Finally, 
note that very similar problems for structures of finite length were considered 
in studies 
\cite{luongo2001mode,
abramyan2011oscillations,
abramian2014oscillations,
abramian2017oscillations}.

\section{Mathematical formulation}
\label{MODE:sec-state-SPRING}
\begin{figure}[htpb]
\centering\includegraphics[width=0.45\textwidth]{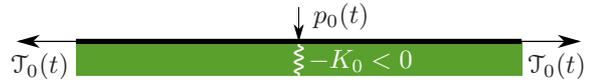}
\caption{The schematic of the system}
\label{winkler-spring.eps}
\end{figure}

We consider transverse oscillation of an infinite taut string on the Winkler
elastic foundation. The elastic foundation has a point inhomogeneity in the form of a concentrated spring of a negative stiffness.
\NEW{The schematic of the system is shown in
Figure~\ref{winkler-spring.eps}.}
Introduce the following notation:
$u(x,t)$ is the displacement of a point of the string at the position $x$ and time~$t$, 
$\mathscr T_0(t)>0$ is the string tension (a given function of time),
$\rho>0$ is the mass of the string per unit length,
$K_0$ is the absolute value for the stiffness $-K_0$ of the concentrated
spring, $k_0>0$ 
is the stiffness for 
the Winkler foundation, $P_0(t)$ is the unknown force on the string from
the spring, $p_0(t)$ is the given external force on the string.
Quantities $k_0$, $K_0$,
$\rho$ are constants. 
One can choose any orthogonal to the string direction, where 
$u(x,t),\ p_0(t)$ and $P_0(t)$ are assumed to be positive.

The governing equations
are
\begin{gather}
\mathscr T_0(t)\,u_{xx}-\rho u_{tt}
-k_0u=-P_0(t)\,\delta(x),
\label{maineq-SPRING-dim}\\
P_0(t)=K_0u(0,t)+p_0(t).
\label{force-spr-dim}
\end{gather}
Here $\delta$ is the Dirac delta-function. 

Now we introduce the dimensionless variables 
\begin{equation}
\tau=t\sqrt{k_0/\rho},\qquad
\xi=x\sqrt{k_0/\hat{\mathscr T}_0}
\label{coo-dless}
\end{equation}
and rewrite governing equations~\eqref{maineq-SPRING-dim},
\eqref{force-spr-dim}
in the following form
\begin{gather}
c^2(\tau)\,\prpr u-\ddot u
-u=-P(\tau)\,\delta(\xi),
\label{maineq-SPRING}
\\
P(\tau)=2u(0,\tau)+2p(\tau),
\label{force-spr}\\
\hat{\mathscr T}_0=\frac{K_0^2}{4k_0},\\ 
c^2=\mathscr T_0/\hat{\mathscr T}_0,\quad
P=P_0/\sqrt{k_0\hat{\mathscr T}_0},\quad 2p=p_0/\sqrt{k_0\hat{\mathscr T}_0}
.
\label{P-dless}
\end{gather}
Here and in what follows, we denote by prime the derivative
with respect to spatial coordinate $\xi$ and
by overdot the derivative with respect to
time $\tau$. The quantity $c>0$ is the dimensionless transverse
waves speed. 
\NEW{The quantities $u,\ P,\ p$ remain dimensional ones having the
dimension of length.}

\NEW{%
The initial conditions for Eq.~\eqref{maineq-SPRING} can be
formulated in the following form, which is conventional for distributions (or
generalized functions) \cite{Vladimirov1971}:
\begin{equation}
u \big|_{\tau<0}\equiv0.
\label{<0-SPRING}
\end{equation}
Note that according to 
Eqs.~\eqref{maineq-SPRING},
\eqref{force-spr},
\eqref{<0-SPRING}
we restrict ourselves to the important particular case of the general problem
concerning non-stationary oscillation, where any external excitation (and, in
particular, non-zero initial conditions) is applied only to the point of the
string under the concentrated spring. One can take into account non-zero initial conditions 
\begin{equation}
\begin{aligned}	
&u(0,0)=0,\qquad &&\dot u(0,0)=2v_0;\\
&u(\xi,0)=0,\qquad &&\dot u(\xi,0)=0\qquad (\xi\neq0);
\end{aligned}
\label{ic-v0}
\end{equation}
for the
velocity of the point
under the concentrated spring, 
introducing of $p$ in the form of \cite{Vladimirov1971}
\begin{equation}
p=v_0\delta(\tau)+\bar p(\tau).
\label{p-delta}
\end{equation}
Here $\bar p(\tau)$ is a non-singular at $\tau=0$ function 
such that $\bar p(\tau)\big|_{\tau<0}\equiv0$.}

The problem under consideration 
\eqref{maineq-SPRING},
\eqref{force-spr},
\eqref{<0-SPRING}
is symmetric with respect to $\xi=0$. Integrating 
\eqref{maineq-SPRING}
over $\xi=0$ results in the following condition 
\begin{equation}
[u']=-\frac {P(\tau)}{c^2(\tau)}=-
\frac{2u(0,\tau)+2p(\tau)}{c^2(\tau)}.
\label{1dof-SPRING-p}
\end{equation}
Here, and in what follows, $[\mu]\equiv\mu(\xi+0)-\mu(\xi-0)$ for any arbitrary quantity
$\mu$. Due to symmetry one has $[u']=2u'(\xi+0)$. Thus, the problem for
infinite string can be equivalently reformulated as the problem for
homogeneous equation 
\begin{gather}
c^2(\tau)\,\prpr u-\ddot u
-u=0
\label{coo-dless-semi}
\end{gather}
for $\xi>0$
with  boundary condition at $\xi=0$
\begin{equation}
u'(0,\tau)=
-\frac{u(0,\tau)+p(\tau)}{c^2(\tau)}.
\label{force-spr-semi}
\end{equation}
This equivalent formulation 
\eqref{coo-dless-semi},
\eqref{force-spr-semi},
\eqref{<0-SPRING}
is used for numerical calculations
(Section~\ref{section-numerics}). 

\section{\NEW{A string with constant tension}}
In this section we consider the infinite string with a constant tension, thus
$c=\const$.
\subsection{Spectral problem}
Put $p=0$ and consider the steady-state problem concerning the natural
oscillations of
the system described by Eqs.~\eqref{maineq-SPRING}--\eqref{force-spr}.
Take
\begin{equation}
u=\hat u(\xi)\,\exp(-\I\Omega \tauu).
\label{umega-SPRING}
\end{equation}
Let us show
that 
such a system possesses a mixed spectrum of natural frequencies.
There exists a continuous
spectrum of frequencies, which lies higher than the cut-off (or boundary) frequency:
$|\Omega|\geq1$.
The modes corresponding to the frequencies from the continuous spectrum are
harmonic waves. Trapped modes correspond to the frequencies from the
discrete part of the spectrum, which lies lower than the cut-off frequency:
$0<|\Omega|<1$. We want to demonstrate that for the problem under consideration
the only
one trapped mode can exist. 
Trapped modes are modes with finite energy, therefore, we require
\begin{equation}
\Int\hat u^2\,\d\xi<\infty,\qquad
\Int\hat u'{}^2\,\d\xi<\infty.
\label{finite-energy}
\end{equation}
Now we substitute Eq.~\eqref{umega-SPRING} into Eq.~\eqref{maineq-SPRING}.
This yields 
\begin{equation}
 {\hat u}''
 -A^2(\Omega){\hat u}=
 -\frac{2{\hat u}(0)}{c^2}
 \,\delta(\xi),
 \label{ino-SPRING}
\end{equation}
where
\begin{equation}
A^2(\Omega)=
\frac{
1-\Omega^2}{c^2}.
\label{B-SPRING}
\end{equation}
Here, by definition, we assume that
\begin{equation}
A(\Omega)>0 
\label{A-geq0}
\end{equation}
for $0<\Omega<1$.
The dispersion relation for the operator in the left-hand side 
of~\eqref{ino-SPRING} is 
\begin{equation}
 \omega^2
 +A^2(\Omega)=0,
 \label{dis_relation-SPRING}
\end{equation}
therefore, the wavenumber 
$\omega$ can be expressed as follows:
\begin{equation}
\omega=
\pm \I A(\Omega).
\label{MODE:omega_gamma-SPRING}
\end{equation}
The solution of Eq.~\eqref{ino-SPRING}, which satisfies 
\eqref{finite-energy}, is 
\begin{equation}
 \hat u(\xi)=
 {\hat u}(0)
 \,\frac{\exp(-A(\Omega_0)|\xi|)
 }
 {c^2A(\Omega_0)}.
\label{U=U-SPRING} 
\end{equation}
Calculating the right-hand side of Eq.~\eqref{U=U-SPRING} at $\xi=0$ yields 
the frequency equation
\begin{equation}
 c^2A(\Omega_0)=1,
 \label{Omega_0^2-SPRING}
\end{equation}
where $\Omega_0$ is the trapped mode frequency.
Resolving the frequency equation results in
\begin{equation}
 \Omega_0^2=1-{c}^{-2}.
 \label{Omega_0-SPRING}
\end{equation}
Thus, according to 
\eqref{A-geq0},
\eqref{Omega_0^2-SPRING},
\eqref{Omega_0-SPRING}
a trapped mode exists and is unique if and only if 
\begin{equation}	
c>1.
\label{domain}
\end{equation}
The critical value 
$c=1$
corresponds to possibility of localized buckling of the
string. 

Note that considering initial non-dimensionless problem for equations 
\eqref{maineq-SPRING}, \eqref{force-spr}
one can easily show \cite{gavrilov2017trapped} that the trapped mode exists only if
\begin{equation}
K_0>0,
\label{domain2}
\end{equation}
i.e.\ we deal with a destabilizing spring.

\subsection{Inhomogeneous non-stationary problem}
Put now $p\neq0$. Applying to Eq.~\eqref{maineq-SPRING}--\eqref{force-spr}
the Fourier transform in time $\tau$ results in 
\begin{equation}
 {u}_F''
 -A^2(\Omega){u_F}=
 -\frac2{c^2}\big({u_F}(0,\Omega)+p_F(\Omega)\big)
 \,\delta(\xi),
 \label{ino-SPRING-non}
\end{equation}
where $u_F(0,\Omega),\ p_F(\Omega)$ are the Fourier transforms of
$u(0,\tau)$ and $p(\tau)$, respectively.
Resolving Eq.~\eqref{ino-SPRING-non} with respect to $u_F(0,\Omega)$ and
applying the inverse transform yields
\begin{multline}
u(0,\tau)
=\frac1{2\pi}\Int
\frac{p_Fe^{-\I\Omega \tau}\,\d\Omega}{c\SQRT-1}
\\=-\frac1{2\pi c^2}\Int
\frac{p_F(c\SQRT+1)e^{-\I\Omega \tau}\,\d\Omega}{\Omega^2-\Omega_0^2}.
\label{before-stat-phase}
\end{multline}

At first, consider the simplest case $p=v_0\delta(\tau)$ ($p_F=v_0)$
that corresponds to the initial conditions 
\eqref{ic-v0}.
To estimate
asymptotically the integral in the right-hand side of 
\eqref{before-stat-phase} for large times $\tau$, we use the method of stationary phase
\cite{fedoruk1977,nayfeh1993}.
The principal part of the asymptotics is given by the contributions from poles 
\begin{equation}
\Omega=\pm\Omega_0-\I0
\label{poles}
\end{equation}
of the amplitude
function 
\begin{equation}
\frac{c\SQRT+1}{\Omega^2-\Omega_0^2}.
\end{equation}
The terms $-\I0$ in the right-hand side of 
\eqref{poles} are taken in accordance with principle of limit 	
absorption. Calculating the contributions from the poles 
\cite{fedoruk1977,Non-Stationary}
results in 
\begin{equation}
u(0,\tau)=\frac{2v_0}{\Omega_0c^2}\sin \Omega_0 \tau+O(\tau^{-3/2}), \quad
\tau\to\infty.
\label{undamped-as}
\end{equation}
One can prove \cite{kaplunov1986torsional} that the principal part of the
error term in 
\eqref{undamped-as} is given by the contribution from the boundary frequencies
(the branching points) $\Omega=\pm1$. Thus, for the large times, the
non-stationary response of the system under consideration is undamped
oscillations with the trapped mode frequency $\Omega_0$. 

Now consider more general case when $p(\tau)$ is a vanishing as $\tau\to\infty$ 
function such that its 
Fourier's transform 
$p_F(\Omega)$ does not have singular points 
on the real axis. 
Applying the method of stationary phase to the asymptotic
evaluation of the integral in
the right-hand side of 
\eqref{before-stat-phase} {results in}
\begin{multline}
u(0,\tau)=\frac{2|p_F(\Omega_0)|}{\Omega_0c^2}
\sin\big(\Omega_0 \tau-\arg p_F(\Omega_0)\big)+o(1), \\
\tau\to\infty.
\label{undamped-as-general}
\end{multline}
The asymptotic order of the error term in the last formula depends on 
the properties of $p_F$.
\section{A string with slowly varying tension}
Assume that the string tension $\mathscr T_0$ and, therefore, the dimensionless
transverse wave speed $c$
are slowly varying piecewise monotone
function of the dimensionless time $\tau$:
$c=c(\epsilon \tau)$.
Here $\epsilon$ is a formal small parameter.
We use an approach 
\cite{gavrilov2002etm,Gavrilov-2006-trans,indeitsev2016evolution}
based 
on the modification of the method of multiple scales
\cite{nayfehperturbation}
(Section~7.1.6) for equations with slowly varying coefficients.
The corresponding rigorous proof, which validates such asymptotic approach in
the case of a one degree of freedom system, can be found in
\cite{feschenko1967eng}.
We look for the asymptotics for the solution under the following conditions:
\begin{itemize}
\item 
$
\epsilon=o(1), 
$
\item
$\tauu=O(\epsilon^{-1}),$
\item
$c(\epsilon \tau)$ satisfies restriction 
\eqref{domain} for all $\tau$.
\end{itemize}

To construct the particular solution of 
\eqref{maineq-SPRING}--\eqref{force-spr},
which describes the evolution of the trapped mode of oscillation 
in the case of slowly varying $c$, we  require that in the perturbed system
\begin{itemize}	
\item Frequency equation 
\eqref{Omega_0-SPRING}
for the trapped mode holds for all $\tau$;
\item Dispersion relation 
\eqref{dis_relation-SPRING} at $\xi=\pm0$
holds for all $\tau$.
\end{itemize}
Accordingly, we use the following ansatz ($\tau>0$, $\xi\lessgtr 0$):
\begin{gather}
 u(\xi,\tauu)=W(X,T)\,\exp\varphi(\xi,\tauu),
 \label{slow-repr-SPRING}
 \\
 T=\epsilon \tauu,
 \quad
 X=\epsilon \xi,
 \\
 {\varphi}'=\I\,\omega(X,T),\quad
 \dot{\varphi}=-\I\,\Omega(X,T),
 \label{fast-phases-SPRING}\\
 W(X,T)=\sum_{j=0}^{\infty} \epsilon^{j}{W_j}(X,T).
 \label{W-series-SPRING}
\end{gather}
Here the amplitude $W(X,T)$, the wavenumber $\omega(X,T)$, and the frequency 
$\Omega(X,T)$ are the unknown functions to be defined in accordance with 
Eq.~\eqref{maineq-SPRING}. 
The variables $X,\ T,\ \varphi $ are assumed to be independent. Accordingly,
we use the following representations for the
differential operators:
\begin{equation}
\begin{gathered}
 \dot{(\cdot)}=-
 \I\,\Omega \,\partial_{\varphi }+\epsilon\,\partial_{T},
 \\
 \ddot{(\cdot)}=-\Omega ^2\,\partial^2_{\varphi \varphi }
 -2\epsilon \I\,\Omega \,\partial^2_{\varphi T}
 -\epsilon \I\, {\Omega '}_T\,\partial_{\varphi }
 +O(\epsilon^2),
 \\
 (\cdot)'=
 \I\,\omega\,\partial_{\varphi}+\epsilon\,\partial_{X},\\
 (\cdot)''=-\omega^2\,\partial^2_{\varphi\varphi}
 +2\epsilon \I\,\omega\,\partial^2_{\varphi X}
 +\epsilon \I\, {\omega}'_X\,\partial_{\varphi}
 +O(\epsilon^2).
\end{gathered}
\label{diff-operators}
\end{equation}
We require that $\omega(X,T)$ and $\Omega(X,T)$ satisfy dispersion relation
\eqref{dis_relation-SPRING} and equation 
\begin{equation}
{\Omega}'_X+{\omega}'_T=0
\label{dxx-SPRING}
\end{equation}
that follows from
\eqref{fast-phases-SPRING}.  Since in the case of a string with constant tension the
undamped oscillation can be described by Eq.~\eqref{undamped-as-general},
we assume that
\begin{gather}
\Omega(\pm 0,T)=\Omega_0(T).
\label{Omega-Omega}
\end{gather}
Additionally, we require that 
\begin{gather}
[W]=0,\qquad 
[\varphi]=0.
\end{gather}
In Eq.~\eqref{Omega-Omega}
the right-hand side is defined in accordance with the frequency
equation \eqref{Omega_0^2-SPRING}, wherein $c=c(T)$.
The phase $\varphi(\xi,\tauu)$ should be defined by the formula 
\begin{equation}
 \varphi=\I\int(\omega\,\d\xi-\Omega\,\d\tauu).
\end{equation}

For large times, integrating formally Eq.~\eqref{maineq-SPRING} with respect to $\xi$ over the
infinitesimal vicinity of 
$\xi=0$ taking into account  
\eqref{force-spr},
one gets
\eqref{1dof-SPRING-p}, 
{wherein $p=0$}.
Now we substitute ansatz 
\eqref{slow-repr-SPRING}--\eqref{W-series-SPRING}
and representations \eqref{diff-operators}
into Eq.~\eqref{1dof-SPRING-p} and equate coefficients of like powers $\epsilon$. 
Taking into account frequency equation \eqref{Omega_0^2-SPRING},
and Eq.~\eqref{Omega-Omega},
one obtains that to the first approximation
\begin{equation}
[W_0{}'_X]=0.
\label{1dof-1st-app-SPRING}
\end{equation}

On the other hand, the quantity in the left-hand side of 
\eqref{1dof-1st-app-SPRING} can be defined by consideration of 
Eq.~\eqref{maineq-SPRING} at $\xi=\pm0$. To do this,
we substitute ansatz 
\eqref{slow-repr-SPRING}--\eqref{W-series-SPRING}
and representations \eqref{diff-operators}
into Eq.~\eqref{maineq-SPRING} 
and equate coefficients of like powers $\epsilon$. 
Taking into account dispersion relation \eqref{Omega_0^2-SPRING} 
and Eq.~\eqref{Omega-Omega},
one obtains that to the first approximation
\begin{equation}
 c^2(T)(2\omega\,\W'_X+{\omega}'_X\,\W)+
 2\Omega_0\,\W'_T+{\Omega_0}'_T\,\W=0
\label{1st-app-XT-SPRING}
\end{equation}
at $\xi=\pm0$.
\def\nu{}
Due to \eqref{dxx-SPRING}
one has
\begin{equation}
 {\omega}'_X={\omega}'_\Omega\,\Omega'_X=
 -{\omega}'_\Omega\,{\omega}'_T,
\label{omega-x}
\end{equation}
where the right-hand side should be calculated in accordance with 
Eq.~\eqref{MODE:omega_gamma-SPRING}.
Thus, Eqs.~\eqref{1st-app-XT-SPRING} and \eqref{omega-x} result in 
\begin{gather}
 \W'_X=-\frac{2\nu\Omega_0
 \,\W'_T
+
(-c^2{\omega}'_\Omega{\omega}'_T
 +\nu{\Omega_0}'_T
 )\,\W
}{2\I\gamma A(\nu\Omega_0,T)c^2(T)},
\label{MODE:W'-SPRING}
\end{gather}
where $\gamma=\sign \xi$. Accordingly,
\begin{gather}
{} [\W'_X]=
 -\frac{\Lambda_2 {W}_0
 +\Lambda_1 {W}_0{}'_T}{\I A(\nu\Omega_0)c^2},
 \label{jump:-SPRING}\\
 \displaybreak[0]
 \Lambda_1\equiv2\nu\Omega_0,
\label{Lambda1}
\\
 \displaybreak[0]
 \Lambda_2\equiv
c^2 A'_\Omega(\nu\Omega_0)A'_T(\nu\Omega_0)
+\nu{\Omega_0}'_T,
\label{Lambda2}
\end{gather}
where the right-hand side of Eq.~\eqref{jump:-SPRING}
is taken at $\xi=0$.

Now, equating the right-hand sides of Eqs.~\eqref{1dof-1st-app-SPRING} 
and \eqref{jump:-SPRING} results in the first approximation equation for
$\bar W_0(T) \equiv W_0(0,T)$:
\begin{equation}
\Lambda_2\bar {W}_0+\Lambda_1\bar {W}_0{}'_T=0.
\end{equation}
The general solution of the last equation is 
\begin{equation}
\bar {W}_0=C_0\exp\Big(-\int\frac{
\Lambda_2}{\Lambda_1 }\,\d T  \Big),
\label{sol-K-var-SPRING}
\end{equation}
where $C_0$ is an arbitrary constant.
Using 
\eqref{B-SPRING},
one gets
\begin{gather}
A'_\Omega=-\frac{\Omega_0}{c\sqrt{1-\Omega_0^2}},
\label{AO}
\\
A'_T=-\frac{\Omega_0\Omega_0{}'_T}{c\sqrt{1-\Omega_0^2}}
-\frac{c{}'_T}{c^2}\sqrt{1-\Omega_0^2}.
\label{AT}
\end{gather}
Substituting these expressions into 
\eqref{Lambda2} yields
\begin{gather}
\Lambda_2=\frac{\Omega^2_0{\Omega_0}'_T}{1-{\Omega}^2_0}+\frac{\Omega_0c'_T}{c}+{\Omega_0}'_T.
\end{gather}
Therefore,
\begin{multline}
-\int\frac{ \Lambda_2}{\Lambda_1 }\,\d T
=-\int\frac{\d\Omega^2_0}{4\left(1-{\Omega}^2_0\right)}
-\int\frac{\d c}{2c}
-\int\frac{\d\Omega_0}{2\Omega_0}\\
=\frac{1}{4}\ln\left(1-{\Omega}^2_0
\right)
-\frac{1}{2}\ln c
-\frac{1}{2}\ln\Omega_0.
\label{int-var-T0}
\end{multline}
Substituting the last equation into the right-hand side of Eq.~\eqref{sol-K-var-SPRING}
yields the final result
\begin{equation}
\bar {W}_0
=C_0\frac{\left(
1-{\Omega}^2_0\right)^{1/4}}{\left(c\Omega_0\right)^{1/2}}.
\label{sol-final-SPRING-var-T0}
\end{equation}
Taking into account 
\eqref{Omega_0-SPRING}, one can rewrite the last formula in two equivalent
forms:
\begin{equation}
\bar {W}_0
=\frac{C_0}{\sqrt c\,(c^2-1)^{1/4}}
=C_0\sqrt{\frac{1-\Omega_0^2}{\Omega_0}}
\label{sol-final-SPRING-var-T0-mod}
\end{equation}
If $\Omega_0 \to +0$ (or, equivalently, $c\to1+0$,
$\mathscr T_0 \to \frac{K_0^2}{4k_0}+0$), then
\begin{equation}
\bar {W}_0 = \frac{C_0}{\Omega_0^{1/2}}+o(1).
\label{A-propto-0}
\end{equation}
Hence, localized low-frequency oscillations
with increasing amplitude precede the
localized string buckling. 

This result is analogous to the
classical result for a one degree of freedom system 
\begin{equation}
\ddot y+\hat\varOmega^2(\epsilon \tau) y =0, 
\label{1dof}
\end{equation}
where the following formula 
\begin{equation}
Y\propto \frac{1}{\hat\varOmega^{1/2}}
\end{equation}
for the amplitude of free oscillations $Y$ is valid 
(the Liouville -- Green approximation \cite{nayfehperturbation}). 
On the other hand, unlike one degree of freedom system \eqref{1dof}, for
the system under consideration, formula 
\eqref{A-propto-0}
is valid only in the limiting case $\Omega_0 \to +0$. 
For finite $\Omega_0$ the dependence 
\eqref{sol-final-SPRING-var-T0-mod}
is more complicated.

\label{MODE:SEC:init-c-SPRING}

Combining the solution in the form of 
Eqs.~\eqref{slow-repr-SPRING}--\eqref{W-series-SPRING}
with its complex conjugate,
we get the non-stationary solution as the following ansatz:
\begin{equation}
u(0,\tau)\sim
\bar W_0\big(\Omega_0(T)\big)\sin\left(\int_0^\tau\Omega_0(T)\,\d T-D_0\right),
\label{anzatz-final}
\end{equation}
where $\bar W_0$ is defined by 
\eqref{sol-final-SPRING-var-T0-mod}.
The unknown constants $C_0$ and $D_0$ should be defined by equating the
right-hand sides of 
\eqref{undamped-as-general} and 
\eqref{anzatz-final} taken at $\tau=0$. This yields
\begin{multline}	
C_0=\frac{2(c^2(0)-1)^{1/4}}{\Omega_0(0)c^{3/2}(0)}\,\big|p_F\big(\Omega_0(0)\big)\big|
\\=\frac{2\big|p_F\big(\Omega_0(0)\big)\big|}{c(0)(1-c^{-2}(0))^{1/4}},
\label{C0-an}
\end{multline}
\begin{equation}	
D_0=\arg p_F\big(\Omega_0(0)\big).
\label{D0-an}
\end{equation}
In the particular case 
$p=v_0\delta(\tau)$
that corresponds to the initial conditions 
\eqref{ic-v0}
one has
\begin{gather}	
C_0=\frac{2v_0\big(c^2(0)-1\big)^{1/4}}{\Omega_0(0)c^{3/2}(0)}
=\frac{2v_0}{c(0)(1-c^{-2}(0))^{1/4}},
\label{C0-an-simple}
\\
D_0=0.
\label{D0-an-simple}
\end{gather}
In the particular case  
\begin{equation}
p(\tau)=\bar p_0H(\tau)\exp(-\lambda \tau),
\label{MASS:f-exp}
\end{equation}
where 
$H(\tau)$ is the Heaviside function, $\bar p_0=\const$, and 
$\lambda=\mathrm{const}>0$, one has
\begin{gather}
p_F\big(\Omega_0(0)\big)=\frac{\bar p_0}{
\lambda-\I\Omega_0(0)},
\label{psi-exp}
\\
\displaybreak[0]
\big|p_F\big(\Omega_0(0)\big)\big|=\frac{|\bar p_0|}{\sqrt{\lambda^2+\Omega_0(0)^2}},
\label{psi-exp-abs}
\\
\displaybreak[0]
\arg\big(p_F\big(\Omega_0(0)\big)\big)=\arctan \frac{\Omega_0(0)}\lambda.
\label{psi-exp-arg}
\end{gather}%
Finally, using
in the case $p=v_0\delta(\tau)$ approximate asymptotically incorrect formula 
\eqref{A-propto-0} 
instead of correct formula
\eqref{sol-final-SPRING-var-T0-mod}
yields
\begin{equation}
C_0=\frac{2v_0}{\sqrt{\Omega_0(0)}\,c^2(0)}=\frac{2v_0}{{\big(1-c^{-2}(0)\big)^{1/4}}\,{c^2(0)}}
\label{C0-an-wrong}
\end{equation}
together with
formula \eqref{D0-an-simple}.

\section{Numerics}
\label{section-numerics}
In previous studies
\cite{gavrilov2002etm,Gavrilov-2006-trans,indeitsev2016evolution,gavrilov2017trapped} we dealt
with several problems for the linear Klein-Gordon equation with constant
coefficients. Numerical solutions were obtained by means of the reduction of
the corresponding problem to an integral Volterra equation of the second kind 
with its kernel
expressed in terms of the fundamental solution of the Klein-Gordon equation.
This cannot be done for the problem under consideration in this paper, since
now we deal with an equation with time-varying coefficients. To perform the numerical
calculations we use 
{\sc SciPy} software.

To verify constructed analytical solution 
\eqref{sol-final-SPRING-var-T0} we solve numerically the
initial value problem 
for PDE
\eqref{coo-dless-semi}
with boundary condition \eqref{force-spr-semi}
using the finite difference method. 
To discretize PDE 
\eqref{coo-dless-semi}
we use the following implicit difference scheme:
\begin{multline}
(c^i)^2
\frac{u_{j+1}^{i}-2u_j^{i}+u_{j-1}^{i}}{(\Delta \xi)^2}\\-
\frac{u_j^{i+1}-2u_j^{i}+u_j^{i-1}}{(\Delta \tau)^2}-
\frac{u_j^{i+1}+u_j^{i-1}}2=0,
\end{multline}
where integers $i,\ j$ ($0\leq j\leq N,\ -1\leq i$) are such that
\begin{gather}
u_j^i=u(j\Delta\xi,i\Delta\tau),\\
c^i=c(i\Delta\tau).
\end{gather}
This scheme conserves 
\cite{donninger2011numerical,strauss1978numerical}
the discrete energy for a
nonlinear Klein-Gordon equation with constant coefficients. 
Numeric boundary conditions that correspond to
\eqref{force-spr-semi} are taken in the form \cite{strikwerda2004finite}
\begin{multline}
(c^{i+1})^2\frac{-3u_0^{i+1}+4u_1^{i+1}-u_2^{i+1}}{2\Delta\xi}
\\+
(c^{i-1})^2\frac{-3u_0^{i-1}+4u_1^{i-1}-u_2^{i-1}}{2\Delta\xi}
\\+
{(u_0^{i+1}+u_0^{i-1})}
+
(p^{i+1}+p^{i-1})=0.
\end{multline}
At the right end we use boundary condition
\cite{trangenstein2009numerical}
\begin{equation}
u^i_{N}=u^i_{N-1}
\end{equation}
\NEW{that corresponds to the physical boundary condition
$u'=0$. Actually, the specific form of this boundary condition
is not very important in our calculations, since we consider the discrete model of
the string with sufficiently large length such that the wave reflections at the right
end do not occur.}

Numerical initial conditions are 
\begin{equation}
u^0_j=u^{-1}_j=0.
\end{equation}
All numerical results below are obtained for the case 
\begin{equation}
\Delta\xi=0.025,\qquad 
\Delta\tau=0.01. 
\label{Deltas}
\end{equation}
The dimensional coefficients $v_0$ and $\bar p_0$ are taken as $1\,\mathrm{m}$
and skipped in what follows for the aim of simplicity.
Calculating the numerical solutions, which corresponds to $p=\delta(\tau)$, we
approximate the Dirac delta-function as follows: 
\begin{equation}
p=\tau_0^{-1}(H(\tau)-H(\tau-\tau_0)).
\end{equation}

\begin{figure}[htpb]
\centering\includegraphics[width=0.5\textwidth]{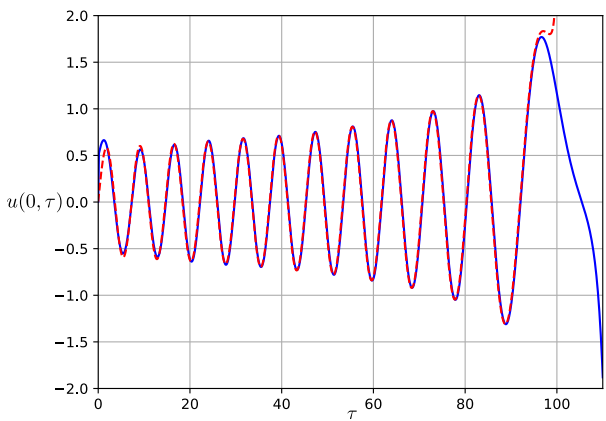}
\caption{
Comparing the analytical solution 
\eqref{sol-final-SPRING-var-T0-mod},
\eqref{anzatz-final},
\eqref{C0-an-simple},
\eqref{D0-an-simple}
obtained for $p=\delta(\tau)$ 
(the red dashed line) and the numerical solution obtained for 
$p(\tau)=\tau_0^{-1}(H(\tau)-H(\tau-\tau_0))$
(the blue solid line) in the case 
$c^2(\epsilon \tau)={4-\epsilon \tau}$. Here $\epsilon=0.03,\ \tau_0=0.1$.
The localized
buckling occurs at $\tau=100$.}
\label{T-kg-simple.eps}
\end{figure}

\begin{figure}[htpb]
\centering\includegraphics[width=0.5\textwidth]{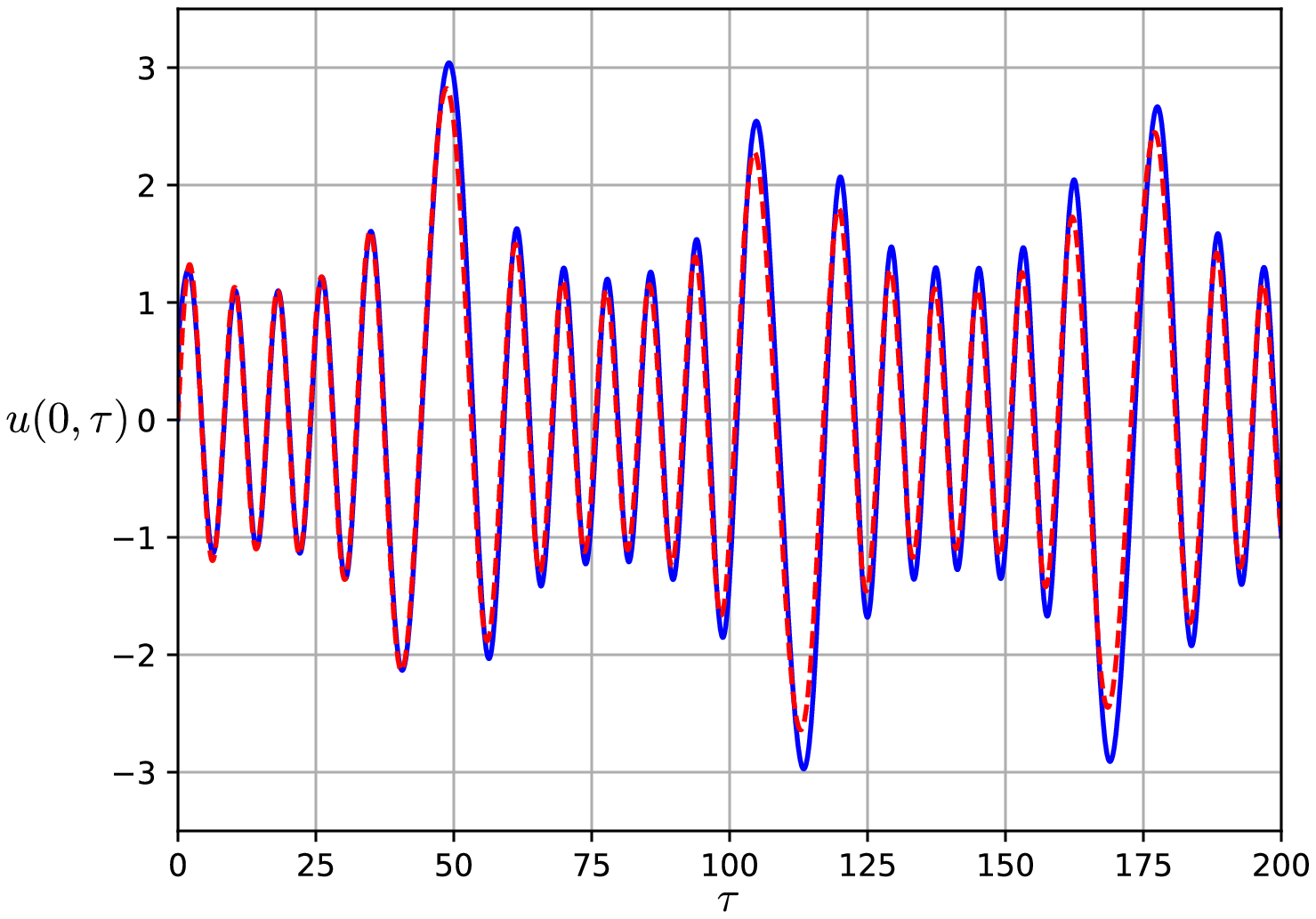}
\caption{
Comparing the analytical solution 
\eqref{sol-final-SPRING-var-T0-mod},
\eqref{anzatz-final},
\eqref{C0-an-simple},
\eqref{D0-an-simple}
obtained for $p=\delta(\tau)$ 
(the red dashed line) and the numerical solution obtained for 
$p=\tau_0^{-1}(H(\tau)-H(\tau-\tau_0))$
(the blue solid line) in the case 
$c^2(\epsilon \tau)=2+0.9\sin(\epsilon \tau)$. Here $\epsilon=0.1,\
\tau_0=0.1.$}
\label{T-kg-osc.eps}
\end{figure}

\begin{figure}[htpb]
\centering\includegraphics[width=0.5\textwidth]{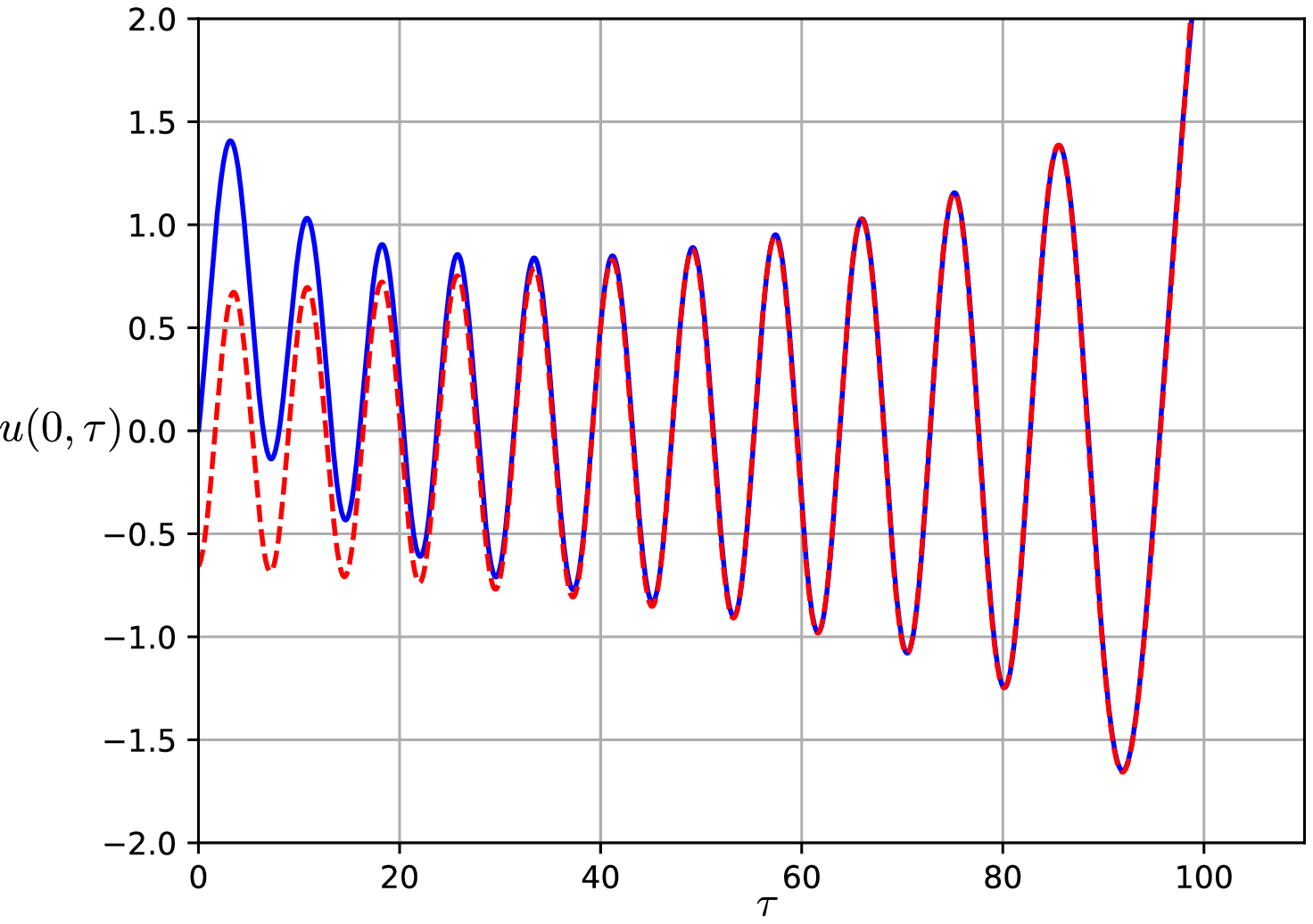}
\caption{
Comparing the analytical solution 
\eqref{sol-final-SPRING-var-T0-mod},
\eqref{anzatz-final},
\eqref{C0-an},
\eqref{D0-an},
\eqref{psi-exp-abs},
\eqref{psi-exp-arg}
(the red dashed line) and the numerical solution obtained for 
$p=H(\tau)\exp(-\lambda\tau)$ 
(the blue solid line) in the case 
$c^2(\epsilon \tau)={4-\epsilon \tau}$. Here $\epsilon=0.03,\ \lambda=0.1$.
The localized
buckling occurs at $\tau=100$.}
\label{T-kg-lexp.eps}
\end{figure}

\begin{figure}[htpb]
\centering\includegraphics[width=0.5\textwidth]{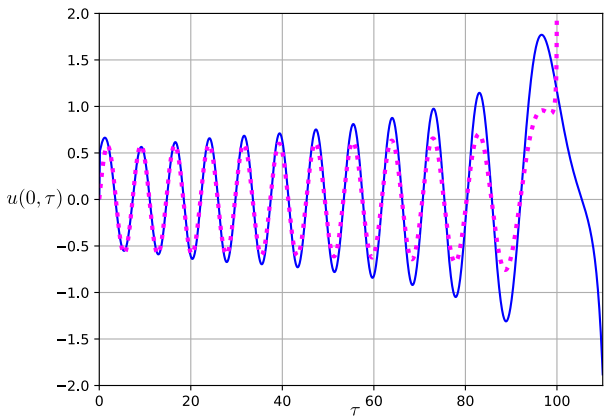}
\caption{
Comparing the approximate asymptotically incorrect analytical solution 
\eqref{A-propto-0},
\eqref{anzatz-final},
\eqref{C0-an-wrong},
\eqref{D0-an-simple}
obtained for $p=\delta(\tau)$ 
(the magenta dotted line) and the numerical solution obtained for 
$p=\tau_0^{-1}(H(\tau)-H(\tau-\tau_0))$
(the blue solid line) in the case 
$c^2(\epsilon \tau)={4-\epsilon \tau}$. Here $\epsilon=0.03,\ \tau_0=0.1$.
The localized
buckling occurs at $\tau=100$.}
\label{T-kg-simple-wrong.eps}
\end{figure}

A comparison between the analytical and numerical solutions is presented
in Figures~\ref{T-kg-simple.eps}--\ref{T-kg-simple-wrong.eps}.
In Figure~\ref{T-kg-simple.eps} we compare the results obtained for the case of
$p=\delta(\tau)$ and monotonically decreasing $c(\epsilon \tau)$. The asymptotic
solution approaches the numeric one very quickly.
The localized
buckling occurs at $\tau=100$ that corresponds to the critical value $c=1$.
In Figure~\ref{T-kg-osc.eps} we compare the results obtained for the case of
$p=\delta(\tau)$ and slowly oscillating $c(\epsilon \tau)>1$.
In Figure~\ref{T-kg-lexp.eps} we compare the results obtained for the case of
$p(\tau)=H(\tau)\exp(-\lambda\tau)$ and monotonically decreasing $c(\epsilon
\tau)$. Since $\lambda=0.1$ is taken small enough, the method of the stationary phase 
gives a reasonable result only after some time ($\tau\approx30$). After that
time the analytical solution approaches the numerical one. 
Finally, in Figure~\ref{T-kg-simple-wrong.eps} we compare the results 
obtained for the case of $p=\delta(\tau)$ and monotonically decreasing
$c(\epsilon \tau)$, using approximate asymptotically incorrect formula 
\eqref{A-propto-0} that corresponds to the Liouville -- Green approximation
for a one degree of freedom system 
\eqref{1dof}.
One can see that in the last case the analytical solution and the numerical one
diverge.


\section{Conclusion}

In the paper we consider a non-stationary localized oscillations of an
infinite string
with slowly time-varying tension. The string lies on the
Winkler foundation with a point elastic inhomogeneity (a concentrated spring
with negative stiffness). 
\NEW{We restrict ourselves to the important particular case of the general problem
concerning non-stationary oscillation, where any external excitation (and, in
particular, non-zero initial conditions) is applied only to the point of the
string under the concentrated spring.}
In the case of the string with a constant tension a 
trapped mode of oscillations exists and is unique if and only if conditions 
\eqref{domain}, \eqref{domain2} are satisfied. The existence of a trapped mode
leads to the possibility of the wave localization near the inhomogeneity 
in the Winkler foundation.
Applying a vanishing as $\tau\to\infty$ external excitation to the
point of the string under the inclusion leads to the emergence of undamped
free oscillations localized near the spring.
The most important result of the paper is analytical formulas 
\eqref{sol-final-SPRING-var-T0-mod},
\eqref{anzatz-final},
\eqref{C0-an},
\eqref{D0-an},
which allow us to describe for large times such non-stationary localized
oscillations in the case of slowly time-varying string tension. The obtained
analytical results were verified by independent numerical calculations
based on the finite difference method. 
The applicability of the analytical formulas was demonstrated for various types
of external excitation and laws governing the varying tension
(see Figures \ref{T-kg-simple.eps}--\ref{T-kg-lexp.eps}).

We also have shown that localized low-frequency oscillations with 
increasing amplitude precede the localized string buckling (see 
\eqref{A-propto-0}). However, unlike the case \eqref{1dof} of a one degree of freedom
system with time-varying stiffness,
in the framework of the problem under consideration
formula \eqref{A-propto-0} is correct only in
the limiting case, where the frequency of localized oscillations approaches zero. 
This fact is also verified by independent numerical calculations
(see Figures \ref{T-kg-simple.eps}, 
\ref{T-kg-simple-wrong.eps}).

\NEW{Finally, we note that in order to consider a more general problem, where the
external excitation can be applied to an arbitrary point of the string, we
need to take into account wave reflections and passing through the discrete
inclusion. This makes the problem to be more  sophisticated. It may be a subject
of a separate future study.}

\begin{acknowledgements}
The authors are grateful to Prof. D.A.~Indeitsev for useful and stimulating discussions.
\end{acknowledgements}

\section*{Compliance with ethical standards}
Conflict of Interest:
The authors declare that they have no conflict of interest.


\end{document}